\documentclass[runningheads]{llncs}
\usepackage{graphicx}
\usepackage{todonotes}
\usepackage{hyperref}

\newcommand{\axis}{\textsf{Axis}}

\usepackage[margin=0.75in]{geometry}

\usepackage{pgfplots}
\usetikzlibrary{patterns}

\begin{document}

\title{Evaluating the Future Device Security Risk Indicator\\ for Hundreds of IoT Devices}
\titlerunning{Evaluating the FDSRI for hundreds of IoT Devices}
\author{Pascal Oser\inst{1,2}\orcidID{0000-0002-2461-343X} \and
Felix Engelmann\inst{3}\orcidID{0000-0001-9356-0231} \and\\
Stefan Lüders\inst{1}\orcidID{0000-0001-8676-2353} \and
Frank Kargl\inst{2}\orcidID{0000-0003-3800-8369}}

\authorrunning{P. Oser et al.}
\institute{European Organization for Nuclear Research (CERN), Geneva, Switzerland \newline
\email{stefan.lueders@cern.ch}\\
\and
Ulm University, Ulm, Germany\\
\email{\{pascal.oser, frank.kargl\}@uni-ulm.de}
\and
IT University of Copenhagen\\
\email{fe-research@nlogn.org}}
\maketitle              
\begin{abstract}

IoT devices are present in many, especially corporate and sensitive, networks and regularly introduce security risks due to slow vendor responses to vulnerabilities and high difficulty of patching.
In this paper, we want to evaluate to what extent the development of future risk of IoT devices due to new and unpatched vulnerabilities can be predicted based on historic information.
For this analysis, we build on existing prediction algorithms available in the SAFER framework (prophet and ARIMA) which we evaluate by means of a large data-set of vulnerabilities and patches from 793 IoT devices.
Our analysis shows that the SAFER framework can predict a correct future risk for 91\,\% of the devices, demonstrating its applicability.
We conclude that this approach is a reliable means for network operators to efficiently detect and act on risks emanating from IoT devices in their networks.

\end{abstract}
%
%
\section{Introduction} \label{Introduction}

IoT devices are becoming more wide-spread in areas such as smart homes, smart cities, but also in research and office environments. Their sheer number, heterogeneity and limited patch availability provide significant challenges for the security of local networks and the internet in general.
This stems from the observation that many devices have vulnerabilities and the availability of patches varies greatly by device and vendor.
The systematic evaluation of device risks, which is essential for mitigation decisions, is currently a skill-intensive task that requires expertise like network vulnerability scanning, or even manual firmware binary analysis.

This paper presents an in-depth and large-scale IoT device security assessment by utilizing the risk \emph{prediction \& scoring} component of the Security Assessment Framework for Embedded-device Risks (SAFER)\cite{oser-tops}. 
SAFER is a highly-automated framework to identify devices on the network for estimating their current as well as future security risk based on publicly retrievable firmware information.
For this work, we focus on the future prediction quality as the current risk of a device is calculated deterministically.
To estimate device risks, we rely on information from among others 
public vulnerability databases, vendor published software license statements and firmware release notes. 
Moreover, based on past vulnerability data and vendor patch intervals for device models, SAFER’s risk component extrapolates those observations into the future using different predefined and automatically parameterized prediction models. This lets SAFER estimate an indicator for future device security risks enabling users to be aware of devices exposing high risks in the future. 
 
One major strength of using SAFER’s risk component over other approaches is the ability to perform significantly automated risk assessments for risks associated to the current firmware, the detection of already patched vulnerabilities and the estimation of a future device risk indicator based on past observed and estimated future information. 

Oser et al.~\cite{oser-tops} describe SAFER extensively, however the evaluation focuses on device identification and the authors only performed a preliminary evaluation of the risk metrics with 38 device models. 
They concluded that for their reduced data-set, current and future risk was predicted with almost 100\,\% accuracy. As this perfect prediction could be caused by a too small data sample, it remains to show the prediction performance in a large scale, realistic setting.

To investigate this strength, we deployed a version of SAFER with an enhanced risk \emph{prediction \& scoring} component in the network of a large multinational organization to systematically assess the security level for hundreds of IoT devices in large-scale networks. 
In this work, we utilized SAFER to estimate the risks of 6,123 different firmware versions for 838 device models. The future security risk is calculated from a patch trend, indicating how long vendor needs to patch vulnerabilities, and a vulnerability trend, indicating the likely severity of future vulnerabilities. For the combined future device security risks, SAFER achieved correct predictions for 91.30\,\% of 793 device models using vendor published information.
This shows that the preliminary evaluation lacked in depth and that SAFER is indeed a valuable tool in realistic settings with many devices and can guide administrators to identify devices that are likely to cause security problems in the future.

\section{Related Work}
\label{section:related_work}

To give an overview of the field, we first introduce related work for the separate aspects of our contribution like vulnerability prediction and risk scoring, as we are not aware of a similar combined work.

\subsubsection{Vulnerability Prediction}

Wu et al.~\cite{wu2020vulnerability} use a multi-variable Long Short-Term Memory (LSTM) to predict time-series data for vulnerabilities. The authors use browser vulnerabilities from May 2008 to
May 2019 of the National Vulnerability Database (NVD) to retrieve vulnerabilities for five web browsers which they use to evaluate their approach on.
An Auto-regressive Integrated Moving Average (ARIMA) model is also trained on this data to compare the results with the proposed solution. The authors state that their solution predicts the number of vulnerabilities for Chrome better than an ARIMA model and achieve an RMSE of 8.032 by their approach and an RMSE of 15.210 by using an ARIMA model. The RMSE is based on the error of the predicted vulnerabilities for 40\,\% of the browser data-set.

ReVeal by Chakraborty et al.~\cite{chakraborty2021deep} performs vulnerability prediction on the Linux Debian Kernel and Chromium. The authors use these code bases because those are well-maintained public projects with large evolutionary history including plenty of publicly available vulnerability reports. They compare ReVeal with four other approaches (\cite{russell2018automated,li2018vuldeepecker,li2021sysevr,zhou2019devign}) of related work and state that existing works have the following limitations: 1) they introduce data duplication, 2) do not handle data imbalance, 3) do not learn semantic information, and 4) lack class separability. The authors compare their Deep-Learning approach with the best performing model in the literature and state that ReVeal performs up to 33.57\,\% better in precision and 128.38\,\% better in recall. 

Dam et al.~\cite{dam2017automatic} propose an LSTM model to learn both semantic and syntactic features of code. With this knowledge, the model predicts vulnerabilities for 18 Android applications written in Java. The authors claim a prediction improvement, compared to traditional software metrics approaches, of 3-58\,\% for within-project prediction and 85\,\% for cross-project prediction. 

Jimenez et al.~\cite{jimenez2016vulnerability} analyzed the Linux Kernel with more than 570,000 commits from 2005 to 2016. They observed that approaches based on header files and function calls perform best for future vulnerability prediction. The authors state that text mining is the best technique when aiming at random instances and identified that code metrics, on the opposite, perform poorly.

\subsubsection{Risk Assessment Approaches}
In the following, we introduce risk assessment approaches partly focused on IoT.
Bahizad~\cite{bahizad2020risks} discusses the increase of IoT devices and the resulting risks a network has to deal with. Bahizad states: ``[...] due to the connections between IoT devices, the security of one device is also dependent on the security of other devices and the cascading effects of its vulnerabilities to the whole system. As these devices increase, the risk added to the system increases.''~\cite{bahizad2020risks}.
This observation emphasizes the problems existing in large heterogeneous networks of IoT devices and motivates the need for IoT risk assessment solutions and their performance evaluation in real world scenarios. 

Shivraj et al.~\cite{shivraj2017graph} propose a model-driven risk assessment framework based on graph theory.  
The authors modeled the system using attack trees and simulated different modes of attack propagation on it. Ultimately, they stress the usefulness of their work by empirical analysis and experiments on the STRIDE\footnote{\url{https://msdn.microsoft.com/en-us/library/ee823878(v=cs.20).aspx}} and LINDDUN\footnote{\url{https://www.linddun.org}} threat models.
The authors performed an in-depth analysis of 207 vulnerabilities to identify the time until a vulnerability was discovered and advisories released. On average, it takes 5.3 years until the investigated vulnerabilities are publicly registered. As a conclusion, they propose guidelines to improve reporting and consistency of ICS CVE information.

Li et al.~\cite{li2021understanding} developed a method to detect security risks of devices based on firmware fingerprinting. The authors retrieved 9,716 firmware images from third-party websites and 347,685 security reports. 
For firmware fingerprinting, they identified subtle differences in firmware file-systems as well as analyzing contained HTML files after labeling them manually. Using word embeddings in combination with two-layer neural networks and regular expressions, the authors claim 91\,\% precision and 90\,\% recall for fingerprinting firmware images. 6,898 security reports contain firmware and vulnerability information. The authors also identified that more than 10\,\% of detected firmware vulnerabilities do not have any patches listed in public databases.

Duan et al.~\cite{duan2021automated} propose an automated security assessment framework for IoT networks. The proposed security model automatically assesses the security of the IoT network by capturing potential attack paths and identifying the most vulnerable ones. They use machine learning and natural language processing to analyze vulnerability descriptions and predict vulnerability metrics of new vulnerabilities with more than 90\,\% accuracy. The predicted vulnerability metrics are used in a graphical security model consisting of attack graphs.

Rodr{\'\i}guez et al.~\cite{rodriguezsuperspreaders} quantified how IoT manufacturers may act as ``superspreaders'' for device infections. The authors scanned the internet during two months for Mirai-infected devices resulting in 31,950 infected IoT devices of 70 unique manufacturers found in 68 countries. 53\,\% of the 70 identified manufacturers offer firmware or software downloads on their websites, 43\,\% provide password changing procedures (as Mirai targets devices using standard passwords), and 26\,\% of the manufacturers offer advice to protect devices from attacks. In total, they identified that nine vendors share almost 50\,\% of the infections identified. 

\subsubsection{Security Scoring Mechanisms}
We introduce approaches of multiple researchers who work on specialized ways of the Common Vulnerability Scoring System (CVSS).
Some propose improved versions of the CVSS~\cite{wang2011improved}, device type specific scoring systems~\cite{qu2016assessing,vilches2018towards}, propose vulnerability assessment methods~\cite{wang2018vulnerability} or risk management for embedded devices~\cite{guillen2014risk}.
Johnson et al.~\cite{johnson2016can} compare the credibility of the CVSS scoring data of different vulnerability sources of NVD, X-Force, OSVDB, CERT-VN, and Cisco.
Le and Hoang~\cite{le2018security} propose an approach to compute the probability distribution of cloud security threats based on Markov chains and the CVSS.

\section{Preliminaries}
Before presenting our evaluation, we first introduce details of the SAFER framework on which our evaluation builds. We also explain what enhancements we have introduced compared to the version described in~\cite{oser-tops}.

\begin{figure*}[h]
  \centering
  \includegraphics[width=0.5\textwidth]{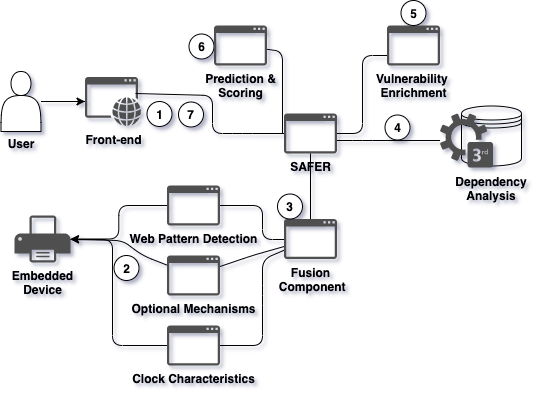}
  \caption{SAFER consists of components to identify devices, fetch their vulnerabilities and score the results.
  }~\label{fig:safer_component_overview}
\end{figure*}
Figure~\ref{fig:safer_component_overview} shows an overview of SAFER and its components.
The SAFER framework implements various components to 1) identify~\cite{oser-CCD,agarwal2019detecting} 
IoT devices, 2) gather vulnerability information for identified devices~\cite{oser-tops}
, 3) score~\cite{oser-tops} 
this information and 4) display~\cite{oser-frontend} 
results to support SAFER's users in making informed decisions regarding their devices' security.

Initiating a device assessment starts by specifying the host-name, IP address or network range which multiple identification mechanisms will scan.
SAFER combines the results of different identification mechanisms using a probabilistic logic framework called Subjective Logic~\cite{josang-subjective-2016}.
The next step is then to retrieve available firmware image and software license statements per firmware version from a vendor's support web-site.
This lets SAFER determine contained software and additional information about the firmware.
Next, the framework gathers publicly known vulnerabilities for identified device models and all contained software using public vulnerability repositories.
In its last step, all information is aggregated and SAFER computes a Current Device Security Risk Indicator (CDSRI) and a Future Device Security Risk Indicator (FDSRI) based on past evidence.
Previous works about SAFER describe its components~\cite{oser-CCD,agarwal2019detecting,oser-frontend} and the overall framework~\cite{oser-tops} but evaluate in particular the FDSRI with only few devices and less details. 

\subsubsection{Predicting a Device's Future Risk}

SAFER extrapolates past observations into a prediction of a future security risk a device may pose due to future vulnerabilities.
It also considers how quickly such future vulnerabilities might be patched by vendors.
To calculate this so called Future Device Security Risk Indicator (FDSRI), SAFER tries to separately predict the frequency and severity level of such future vulnerabilities and also estimates future patch intervals both based on observed historic data.
The FDSRI is composed of two dimensions:
To predict future vulnerability severity levels we calculate a so-called Vulnerability Trend (VT). For this, SAFER takes all previously identified vulnerabilities of a device into account. The Patch Trend (PT), on the other hand, predicts future patch intervals based on time intervals of previously patched vulnerabilities; from the date they became publicly known until the vendor released a patch.
For those predictions, SAFER applies various  prediction models on its vulnerability severity and patch data, specifically Facebook Prophet~\cite{taylor2018forecasting}, simple moving average, an auto regressive model\footnote{\url{https://www.statsmodels.org/dev/generated/statsmodels.tsa.ar_model.AutoReg.html}} and different Auto Regressive Integrated Moving Average (ARIMA) models\footnote{\url{https://www.statsmodels.org/stable/generated/statsmodels.tsa.arima.model.ARIMA.html}} based on the Box-Jenkins method~\cite{box2011time}.

\section{Predicting the Patch Trend}

This section introduces the evaluation of the \emph{Patch Trend}.
First, we introduce how we built our data-set containing dates of past patch intervals for our device models under investigation. Second, based on the device model's past patch behavior, we evaluate different prediction models to measure how well these predict future patch intervals. Third, we discuss the predicted future patch intervals and how they compare with actually observed patch intervals to identify prediction accuracy.

\subsection{Data-set Observation}
\begin{figure}
    \centering
    
    \begin{tikzpicture}
    \begin{axis}[
      ybar,
      xmin=0,
      xmax=6200,
      ymin=0,
      height=4.5cm,
      width=\textwidth,
      ylabel={Vulnerabilities},
      xlabel={Patch Intervals [years]},
      xtick={365,1825,3650,5475},
      xticklabels={1,5,10,15},
    ]
    \addplot+ [hist={data=x,bins=60}]
        file {data/fig1_raw.dat};
        
    \end{axis}
    \end{tikzpicture}
    
    \caption{Patch interval overview in years for patched vulnerabilities.}
    \label{fig:ch9_patchinterval_overview_plot}
\end{figure}
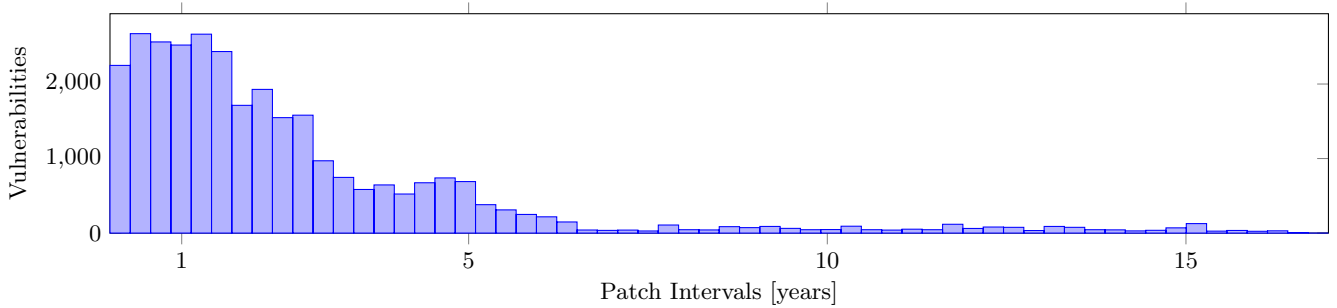

SAFER analyzed, exemplary for \axis{},
hundreds of release notes and license statements. Those support SAFER with detailed information about firmware-contents by simply parsing text files. 
The parser is vendor specific and needs to understand the semantics of patch notes. We noticed that the structure is consistent within vendors but differ slightly between them, requiring adaptation of the parser.
SAFER creates a new data-set per device model containing the release dates of relevant CVEs and the time in days the vendor required to patch those\footnote{The raw data-set will become available at \url{https://safer.network/}}.
The data also varies on the amount of patches the vendor applied to device models.
SAFER identified that 423 device models have not been patched by their vendor compared to 370 device models having received between 2 and 19 patches.
The data shows an average amount of 38.84 patches for all 793 device models and an average of 83.26 patches for the 370 device models having received at least one patch.
The average interval in days the exemplary analyzed vendor \axis{} requires to patch registered vulnerabilities in their analyzed device models is about 953 days. The data divides in the first quartile ranging to 316 days, the second quartile ranging to 634 days and the third quartile ranging to 1,170 days. The maximum time \axis{} required to patch a device model's vulnerability was 6,017 days resulting in circa 16.5 years.
A histogram of the years (x-axis) \axis{} needed for patching vulnerabilities (y-axis) over all their device models is shown in Figure~\ref{fig:ch9_patchinterval_overview_plot}.

After having established our data set, 
we proceed by using the observed patch intervals per device model and estimate future patch intervals.

\subsection{Prediction Model Evaluation}
To evaluate the predicted patch intervals and, hence, derive a patch trend afterwards, we analyzed the data-sets of past patch time-spans for each device model separately.

Each of the data-sets was split in a 66\,\% training- and 34\,\% test-set. We then trained the device model specific prediction models on the training-set of past patch intervals and predicted future patch intervals which we compared with observed values of the test-set. 

While this was also done in the original version of SAFER, we introduce a number of changes in this paper:
originally, the AR model was not evaluated individually and the parameter \textit{changepoint-prior-scale}\footnote{\url{https://facebook.github.io/prophet/docs/trend\_changepoints.html\#automatic-changepoint-detection-in-prophet}} of Facebook's Prophet was set to a static value. In this work, we enhanced SAFER by performing a semi-automated data analysis and automated 1) the AR model's lags and 2) Prophet's \textit{changepoint-prior-scale} parameter to tune parameters per device model data-set.
This improvement enables the prediction model to handle different behavior regimes, e.g. if a vendor suddenly stopped patching a once frequently patched device.  
Per device model, we evaluated the parameter value for \emph{changepoint\_prior\_scale} which adjusts the trend flexibility from rather static (0.01) to highly fluctuating (2.1) trends. We find that adding those two parameterizing approaches increased the prediction accuracy significantly.

We use \textbf{Facebook's Prophet} as one of our prediction models. There, we changed \emph{seasonality\_mode}\footnote{\url{https://facebook.github.io/prophet/docs/multiplicative\_seasonality.html}} to multiplicative which performed better for PT data as seasonality rather grows with data in comparison to being a constant factor where additive models are used. 
In time series, seasonality refers to fluctuations that occur at regular intervals and prevent regular patterns.

For the \textbf{ARIMA} prediction model, we use \emph{pmdarima}\footnote{\url{https://pypi.org/project/pmdarima/}} to perform automated evaluation of the best fitting ARIMA model including others like SARIMAX. The tool evaluates all $p$, $d$, and $q$ parameters which we configured to be in range from 0 to 12 and uses the augmented Dickey–Fuller test to evaluate the most accurate ARIMA model. We consider the parameter range as sufficient for the automated evaluation of device model data, because higher values are likely to overfit the ARIMA model.

The \textbf{auto-regressive} model (AR) is part of the previously discussed ARIMA model but has the $d$ and $q$ parameters set to $0$. For evaluation, we use the AR model of \emph{statsmodels}\footnote{\url{https://www.statsmodels.org/dev/generated/statsmodels.tsa.ar_model.AutoReg.html}} and used the default configuration which performed best. This means that no seasonality and a constant trend is considered. The lags ($p$ parameter) were evaluated individually per data-set.

Finally, we use the \textbf{Simple Moving Average} (SMA) to compare the other prediction mechanisms to a relatively simple approach for forecasting. SMA considers a defined amount of past data-points when forecasting which is known as the so-called window. Calculating the average of past patch intervals in the window and dividing it by the number of patch-intervals leads to the estimated forecast for the next patch-interval. This process is performed recursively until reaching the end of the test-set. The window size was evaluated using the same approach like for ARIMA. 

The prediction model results are shown in Table~\ref{table:scoring_pt_eval_prediction_result_overview_rmse_stuff}. Our enhanced version of SAFER was able to predict patch intervals for 793 device models requiring a minimum amount of two past patch intervals to estimate future patch intervals.
If a device model did receive less than two patches, the prediction models cannot predict future data as no trend is available and considers the PT of the device model as \emph{slow} as in~\cite{oser-tops}.

To compare how accurate the prediction models estimated the observed patch intervals of the test-set, we calculated the Root Mean Square Error (RMSE) and Median Absolute Deviation (MAD). The RMSE is the square root of the average of the square of all errors. The MAD is defined as ``a measure of scale based on the median of the absolute deviations from the median of the distribution.''~\cite{howell2005median}. MAD and RMSE are important indicators when considering both normally and not normally distributed data.

\begin{table}[h]
    \centering
    \caption{Comparison of average patch interval prediction error}
    \begin{minipage}{0.5\textwidth}
    \begin{tabular}{c c c c c}
      Version & Predictor & Devices & RMSE & MAD \\
        \hline
        Our & Prophet & 793 & 443.75 & 119.0 \\
        Our & ARIMA & 793 & 417.14 & 154.15 \\
        Our & AR & 793 & 53.17 & 25.13 \\
        Our & SMA & 793 & 166.52 & 65.08 \\
        \hline
        Original & Prophet & 38 & - & 14.31\\
        Original & ARIMA & 38 & - & 24.12 \\
    \end{tabular}
    \end{minipage}\hfill
    \begin{minipage}{0.44\textwidth}
    \begin{tikzpicture}
    \begin{axis}[
        ybar=0pt,
        bar width=2mm,
        legend style={at={(0.95,0.95)},
          anchor=north east,legend columns=-1},
        ylabel={MAD},
        y label style={at={(0,0.5)}},
        height=3.5cm,
        width=\textwidth,
        xtick=data,
        xticklabels={Prophet,ARIMA,AR,SMA},
        x tick label style={rotate=30,anchor=east},
        ymax=250,
        ymin=0,
        ]
    \addplot[pattern=north east lines,] coordinates {
      (2, 25.13)
      (1,154.15)
      (0,119.0)
      (3,65.08)
    };
    \addplot[] coordinates {
      (0, 24.12)
      (1,14.31)
    };
   
    \legend{Ours, Original}
    \end{axis}
\end{tikzpicture}
\end{minipage}
    \label{table:scoring_pt_eval_prediction_result_overview_rmse_stuff}
\end{table}

Calculating the median
over all AR device model prediction errors results in an RMSE of 53.17 days and MAD of 25.13 days which is the best performing prediction model. We compare our results with the different ARIMA models achieving an RMSE of 417.14 days and a MAD of 154.15 days. Prophet achieved on median an RMSE of 443.75 days and a MAD of 119.0 days. The simple moving average achieves an RMSE of 166.52 days and a MAD of 65.08 days.
Compared to the original SAFER analysis, our best results are still worse which is expected in a larger, realistic data-set. This highlights the importance of our extended evaluation here.

Predicting patch time-spans is not trivial, because the data has a high amount of variation and ranges from a few days up to years. If the trend for all patch intervals would be more stable, we believe that the auto-regressive model would also estimate future patch-intervals more accurately. If the trend fluctuates heavily, Prophet has a higher residual error rate but a better adaptation to the high variation in comparison to other evaluated prediction models
. Moreover, Prophet's multiplicative model better adapts to the heavily varying data in those forecasts but introduces a higher prediction error which significantly increases the RMSE in comparison to the more stable MAD.
Based on the heavily varying data for device models and data sources, we achieved the most accurate prediction results when evaluating the best configuration per prediction model and data-set before forecasting.

To conclude, AR predicted the future patch intervals with an average error of 53.17 days in RMSE and 25.13 days in MAD. AR predicted the PT category for 318 out of 793 device models correctly.
Due to the heavily varying data per device model and data-source, we do not suggest to choose one single prediction model and configuration for PT forecasting as this would worsen the prediction accuracy. For SAFER's PT, achieving the correct PT category based on predicted patch intervals for a device model is the most important task we want to solve. Thus, we discuss how accurate the predicted patch-intervals in days are in comparison to observed patch intervals of the test-set.

\subsection{Prediction Accuracy}
To show the accuracy of our SAFER version for each device model category like CCTV, we grouped the analyzed device models by their category in Table~\ref{tab:ch9_correct_pt_per_cat}.
The table shows that categorizing the PT (in \emph{Fast}, \emph{Medium} and \emph{Slow}) correctly is not trivial due to less available and heavily fluctuating data, ranging from few days to multiple years for a single device model.
We use the same patch trend categories (interval of vendor response) as Oser et al.~\cite{oser-tops}: \emph{Fast}(0 - 22 days, the median time until an exploit exists), \emph{Medium}(23 - 413 days, the median time until full disclosure) and \emph{Slow}($\geq$414 days).

\begin{table}[h]
    \centering
    \caption{Predicted PT accuracy per device category}
    \begin{minipage}{0.5\textwidth}
    \begin{tabular}{c r r r r r}
         Category & \# D.        & Prophet  & ARIMA     & AR        & SMA          \\
         \hline
         CCTV & 720            & 25.97\,\%& 20.41\,\% & 37.50\,\% & 30.41\,\%  \\
         Streaming & 63        & 33.33\,\%& 34.92\,\% & 46.03\,\% & 42.85\,\%  \\
         Switch & 4            & 0.0\,\%  & 0.0\,\%   & 0.0\,\%   & 0.0\,\%     \\
         Speaker & 3           & 0.0\,\%  & 0.0\,\%   & 0.0\,\%   & 0.0\,\%     \\
         Controller & 2        & 0.0\,\%  & 50.0\,\%  & 50.0\,\%  & 0.0\,\%    \\
         IP2Serial & 1         & 100.0\,\%& 100.0\,\% & 100.0\,\% & 100.0\,\% 
    \end{tabular}
    \end{minipage}\hfill
    \begin{minipage}{0.41\textwidth}
    \begin{tikzpicture}
    \begin{axis}[
        ybar=0pt,
        bar width=2mm,
        legend style={at={(0.95,0.95)},
          anchor=north east,legend columns=-1},
        ylabel={accuracy \%},
        y label style={at={(0,0.5)}},
        height=3.2cm,
        width=\textwidth,
        xtick=data,
        xticklabels={Prophet,ARIMA,AR,SMA},
        x tick label style={rotate=30,anchor=east},
        ymax=70,
        ]
    \addplot[] coordinates {
      (0, 25.97)
      (1,20.41)
      (2,37.50)
      (3,30.41)
    };
    \addplot[pattern=north east lines,] coordinates {
      (0,33.33)
      (1,34.92)
      (2,46.03)
      (3,42.85)
    };
   
    \legend{CCTV, Stream}
    \end{axis}
\end{tikzpicture}
\end{minipage}
    \label{tab:ch9_correct_pt_per_cat}
\end{table}

Table~\ref{tab:ch9_correct_pt_per_cat} shows that the predicted PT category does not often align with the observed PT category for all device models.
Below, we discuss our findings for the prediction models to highlight which PT category was estimated and which category was correct. 
318 out of 793 device models were predicted with the correct PT using AR. It wrongly estimated future PT for 52 device models (14\% of all devices on which there was enough data to perform a prediction) with a too high\,/\,low PT category. We identified that 29 device models were estimated with a too high PT and 23 device models were estimated with a too low PT. The lack of historic patch data prevented AR to predict the PT for 423 device models.
We think that PT predictions will improve over time if vendors support devices longer, state when a patch was applied
and when additional data-sources for SAFER are used to combine information. 
We also think that the prediction error can be decreased even further if vendors introduce regular patch-cycles resulting in less fluctuating patch intervals.

In this section, we presened a fine-grained evaluation of our enhanced SAFER patch trend calculation to estimate future patch intervals per device model and prediction model. The resulting patch trend, combining past and estimated future patch intervals, indicates in which time interval a vendor patches vulnerabilities. 
In standard risk assessments, the risk value is calculated by combining the incidence rate and the impact of the risk. For SAFER's device risk assessments, we consider the likelihood as the patch trend calculated above and the impact of the device risk as the vulnerabilities a device is vulnerable to. Hence, the following section focuses on the latter part: 1) considering a device's known vulnerabilities, and 2) estimating the severity of future vulnerabilities to 3) ultimately identify the most common vulnerability severity level per device model. 

\section{Predicting the Vulnerability Trend}
In order to determine the likely severity levels of an IoT device's future vulnerabilities, we calculate the \emph{vulnerability trend}.
Similar to the previous section, we first introduce the data-set containing severity levels of past security vulnerabilities of our device models as it is provided by SAFER's firmware and release note analysis.
Second, based on the device model's past vulnerabilities, we apply different prediction models to evaluate how well they are able to predict  severity levels of future vulnerabilities.
Third, we compare predicted future vulnerability severity levels with observed ones to identify our prediction accuracy.

\subsection{Data-set Observation}
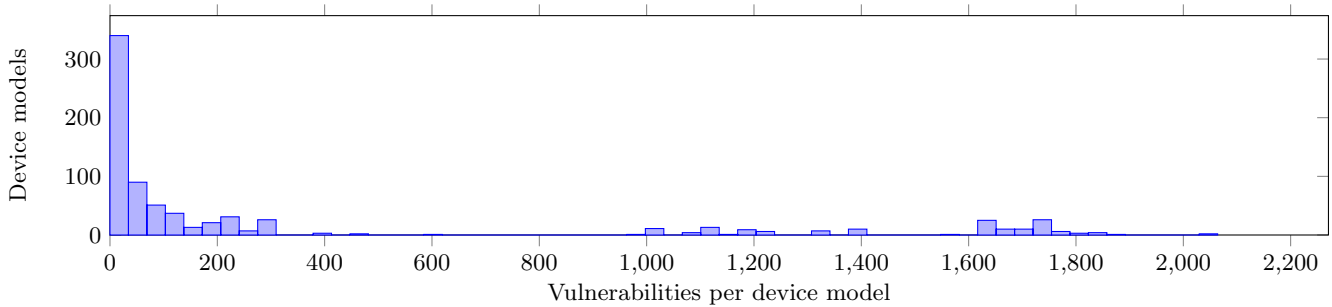
\begin{figure}
    \centering
    \begin{tikzpicture}
    \begin{axis}[
      ybar,
      xmin=0,
      ymin=0,
      height=4.5cm,
      width=\textwidth,
      ylabel={Device models},
      xlabel={Vulnerabilities per device model},
    ]
    \addplot+ [hist={data=x,bins=60}]
        file {data/fig2_raw.dat};
        
    \end{axis}
    \end{tikzpicture}
    \caption{All device models with more than one vulnerability}
    \label{fig:ch9_vulnsofallmodels_overview_plot}
\end{figure}

A histogram of all vulnerabilities for all analyzed device models without duplicates is shown in Figure~\ref{fig:ch9_vulnsofallmodels_overview_plot}.
Our enhanced version of SAFER identified that 293
out of \emph{793} device models
do not have registered vulnerabilities.
This is either based on 1) vendors replacing firmware-contained software so that they do not contain vulnerable software or 2) SAFER did not find a vulnerability in analyzed firmware images. 

Our data further contains 479 device models having between 1 and 2,064 registered vulnerabilities as shown in Figure~\ref{fig:ch9_vulnsofallmodels_overview_plot}, arithmetic average being 549.
Over all device models, the average amount of vulnerabilities is 341.
Based on SAFER's retrieved data, the average vulnerability severity level over all models is 4.9 CVSS. The data is split into the first quartile ranging to 4.4 CVSS, the second quartile ranging to 4.9 CVSS and the third quartile ranging to 7.1 CVSS. Out of all retrieved vulnerabilities, the lowest CVSS severity level is 1.2 CVSS and the highest CVSS is 10.0. 
To estimate future vulnerabilities for device models based on this data-set, we again analyzed different prediction models.

\subsection{Prediction Model Evaluation}
To estimate future vulnerability severity levels for the VT, we use the following prediction models; namely Facebook's Prophet, the auto-regressive integrated moving average (ARIMA) model, the auto-regressive model (AR) and a simple moving average (SMA) model. 
To evaluate the prediction models, we again split our data-set in a 66\,\% training and 34\,\% testing-set per device model. For the VT, the training and test-set contain the severity levels of all vulnerabilities per device model. 
As the VT data is based on vulnerability severity levels, the vertical axis is limited from the minimum 0.0 to the maximum 10.0 CVSS value. The forecast time interval (horizontal axis) is -- analog to PT -- set as the time interval of the test-set.

Our enhanced SAFER version was able to predict future vulnerability severity levels with Prophet for 793, with ARIMA for 770, with AR for 793 and with SMA for 793 out of 793 device models. Based on the correctly estimated device models and lowest prediction error, we consider AR as the best performing prediction model.
SAFER requires a minimum amount of two past vulnerabilities to estimate a trend. If a device's firmware content has less than two vulnerabilities registered, the prediction models cannot predict future data accurately as there is no trend available. This happened for 394 device models for which we did not include a prediction error in Table~\ref{table:scoring_vt_eval_prediction_result_overview}. We argue that for those device models, the vendor actively takes care of replacing vulnerable software which reduces the VT data-set per device model and is represented in SAFER by indicating a \emph{low} VT~\cite{oser-tops}.

We identified the CVSS category of future vulnerabilities for all 793 device models correctly and achieved over all models an RMSE of 1.49 CVSS and a MAD of 0.98 CVSS using AR. This means that we can predict the severity for future vulnerabilities better than a simple moving average or Prophet which both achieve the same amount of correct device models but face a higher prediction error. Since the severity levels of the VT data range from 0.0 to 10.0 there is less fluctuating data. Thus, we use AR -- being a linear model with less variation -- to predict CVSS values from this data source. 

We chose the best performing prediction model AR and discuss the predicted CVSS values hereafter.

\subsection{Prediction Accuracy}
In this section, we compare how accurate our SAFER version estimates the vulnerability trend in comparison to the observed test-data we use to verify the results.

\begin{table}[h]
    \centering
    \caption{VT prediction accuracy overview}
    \begin{minipage}{0.5\textwidth}
    \begin{tabular}{c c c c c c c}
        Version & Model & Corr. VT & Corr. \# & RMSE & MAD \\
        \hline
        Our & Prophet & 100.0\,\% & 793/793 & 2.08 & 1.40 \\
        Our & ARIMA & 97.10\,\% & 770/793 & 1.94 & 1.04 \\
        Our & AR & 100.0\,\% & 793/793 & 1.49 & 0.98 \\
        Our & SMA & 100.0\,\% & 793/793 & 1.74 & 1.19 \\
        \hline
        Orig. & Prophet & 100.0\,\% & 38/38 & - & 2.39 \\
        Orig. & ARIMA & 100.0\,\% & 38/38 & - & 1.31 \\
    \end{tabular}
    \end{minipage}\hfill
    \begin{minipage}{0.41\textwidth}
    \begin{tikzpicture}
    \begin{axis}[
        ybar=0pt,
        bar width=1.5mm,
        legend style={at={(1.00,1.05)},
          anchor=south east,legend columns=2},
        ylabel={CVSS},
        y label style={at={(0,0.5)}},
        height=3.2cm,
        width=\textwidth,
        xtick=data,
        xticklabels={Prophet,ARIMA,AR,SMA},
        x tick label style={rotate=30,anchor=east},
        ymax=2.5,
        ymin=0,
        xmin=-0.8,
        ]
    \addplot[pattern=north east lines,] coordinates {
      (0, 2.08)
      (1,1.94)
      (2,1.49)
      (3, 1.74)
    };
    \addplot[pattern=horizontal lines,] coordinates {
      (0,1.40)
      (1,1.04)
      (2,0.98)
      (3,1.19)
    };
    \addplot[] coordinates {
      (0,2.39)
      (1,1.31)
    };
   
    \legend{RMSE, MAD, original MAD}
    \end{axis}
\end{tikzpicture}
\end{minipage}
    \label{table:scoring_vt_eval_prediction_result_overview}
\end{table}

As Table~\ref{table:scoring_vt_eval_prediction_result_overview} shows, three out of four prediction models estimated the VT correct for all devices.
Only AR estimated the VT category for devices correct with the lowest RMSE of 1.49 CVSS and a MAD of 0.98 CVSS. Thus, our enhanced SAFER version reduces the prediction error in comparison to the original SAFER by 25.2\,\%.
To provide users of SAFER an intuitive understanding of the vulnerability trend, we categorize the most likely vulnerability severity by using CVSS version 2.0 into: \emph{Low} (0.0 - 3.9), \emph{Medium} (4.0 - 6.9) and \emph{High} (7.0 - 10.0).

\begin{table}[h]
    \centering
    \caption{Predicted VT accuracy per device category}
    \begin{tabular}{c c c c c c}
        Category & Devices & AR & ARIMA & SMA & Prophet \\
        \hline
        CCTV & 720 & 100.0\,\% & 95.69\,\% & 100.0\,\% & 100.0\,\% \\
        Streaming & 63 & 100.0\,\% & 100.0\,\% & 100.0\,\% & 100.0\,\% \\
        Switch & 4 & 100.0\,\% & 100.0\,\% & 100.0\,\% & 100.0\,\% \\
        Speaker & 3 & 100.0\,\% & 100.0\,\% & 100.0\,\% & 100.0\,\% \\
        Controller & 2 & 100.0\,\% & 100.0\,\% & 100.0\,\% & 100.0\,\% \\
        IP2Serial & 1 & 100.0\,\% & 100.0\,\% & 100.0\,\% & 100.0\,\% \\
    \end{tabular}
    \label{tab:ch9_correct_vt_per_cat}
\end{table}

Table~\ref{tab:ch9_correct_vt_per_cat} shows that the VT category per device model was often predicted correctly by the prediction models. Apart from the ARIMA prediction model, all three others predicted the VT category correctly for 793 out of 793 device models. However, even though the prediction models achieved 100\,\% for the VT category, they differ in their prediction errors measured in CVSS. 
We conclude that the prediction models of SAFER's enhanced version estimate the future vulnerability severity levels with 25.2\,\% less error.

This section presented the evaluation of our enhanced SAFER version to estimate future vulnerability severity levels per device and prediction model. The resulting VT, combining past and estimated future vulnerability severity levels, indicates the most likely vulnerability severity per device model. 
Combining 1) the most likely time a vendor requires to issue a firmware update (by PT) with 2) the most likely vulnerability severity level per device model (by VT) results in the Future Device Security Risk Indicator (FDSRI) described below.

\section{Predicting the Future Device Security Risk Indicator}
Besides knowing the current risk, it is also relevant for IoT device owners to have an estimate of the risk a device might pose in the future. 
This section refers to the FDSRI introduced in SAFER's original work~\cite{oser-tops} which is based on the vulnerability and patch trend, both containing past observations and future predictions for device models.
We use the patch trend and vulnerability trend in a heuristic to calculate the FDSRI. Then, we discuss how accurate the results for the FDSRI are compared to real, observed data. 

\subsection{Definition of the Future Device Security Risk Indicator} \label{ch9:fdsri_definition}
\begin{table}[h]
\caption{Future Device Security Indicators}
\label{Table:futureSecurityRiskLevel}
\centering
\begin{tabular}{|c||c|c|c|c|}
 \hline & 
\multicolumn{1}{c}{} & 
\multicolumn{1}{|c}{} & 
\multicolumn{1}{c}{  \textbf{Patch Trend}} &
\multicolumn{1}{c|}{} \\
 \hline \hline
  &  &   \textbf{Fast} &   \textbf{Medium} &   \textbf{Slow}\\\cline{2-5}
    \textbf{Vulnerability} &   \textbf{Low} &   Low &   Low &   Medium\\
    \textbf{Trend} &   \textbf{Medium} &   Low &   Medium &   High\\
  &    \textbf{High} &   Medium &   High &   Critical\\
 \hline
\end{tabular}
\vspace{-1ex}
\end{table}

To determine the FDSRI per device model, SAFER needs to combine the device model's previously calculated PT and VT. We use the risk matrix (Table~\ref{Table:futureSecurityRiskLevel}) introduced in SAFER's original work~\cite{oser-tops} to combine the PT and VT in a heuristic leading to the FDSRI. 
We evaluated our approach by first calculating the VT and PT on observed data including the FDSRI. Afterwards, SAFER predicted the patch intervals and future vulnerability severity levels for the test-set which we used in previous sections to calculate the PT and VT on. 
Ultimately, we calculated the FDSRI using both trend (PT and VT) predictions and compare it with the FDSRI derived from real observations. This allows us to verify the accuracy of SAFER's estimated FDSRI.

We highlight that our enhanced SAFER version estimates the correct PT for 40.10\,\% and the correct VT for 100\,\% of 793 device models. We achieve this by individually parameterizing and combining different prediction models to best fit the patch and vulnerability data.

\subsection{Classifying the Future Device Security Risk Indicator} \label{ch9:fdsri_classification}

\begin{table}[h]
    \centering
    \caption{FDSRI accuracy comparison.}
    \begin{minipage}{0.58\textwidth}
    \begin{tabular}{c c c c}
        Version & PT/VT Model & Corr. FDSRI & Corr. \# \\
        \hline
        Our & Prophet & 79.70\,\% & 632/793\\
        Our & ARIMA & 73.52\,\% & 583/793 \\
        Our & AR & 91.30\,\% & 724/793 \\
        Our & SMA & 84.49\,\% & 670/793 \\
        \hline
        Original & Prophet & 100\,\% & 38/38\\
        Original & ARIMA & 100\,\% & 38/38
    \end{tabular}
    \end{minipage}\hfill
    \begin{minipage}{0.41\textwidth}
    \begin{tikzpicture}
    \begin{axis}[
        ybar=0pt,
        bar width=2mm,
        legend style={at={(0.95,0.95)},
          anchor=north east,legend columns=-1},
        ylabel={Corr. FDSRI \%},
        y label style={at={(0,0.5)}},
        ytick={0,50,100},
        height=3.5cm,
        width=\textwidth,
        xtick=data,
        xticklabels={Prophet,ARIMA,AR,SMA},
        x tick label style={rotate=30,anchor=east},
        ymax=160,
        ymin=0,
        ]
    \addplot[pattern=north east lines,] coordinates {
      (0, 79.70)
      (1,73.52)
      (2,91.30)
      (3,84.49)
    };
    \addplot[] coordinates {
      (0, 100)
      (1,100)
    };
   
    \legend{Ours, Original}
    \end{axis}
\end{tikzpicture}
\end{minipage}
    \label{table:scoring_fdsri_eval_prediction_result_overview}
\end{table}

Table~\ref{table:scoring_fdsri_eval_prediction_result_overview} shows how many future device security risk indicators (FDSRI) were identified correctly. In previous sections for VT and PT, we identified that the AR prediction model performed best for the data-source which Table~\ref{table:scoring_fdsri_eval_prediction_result_overview} shows by achieving a correct FDSRI for 91.30\,\%, i.e.\ 724 of 793 device models.

Analogous to VT and PT separately, the original SAFER evaluation with a small sample of devices had an unrealistically high accuracy. Our evaluation therefore provides a realistic prediction performance.

\begin{table}[h]
    \centering
    \caption{Correctly estimated FDSRI per device category}
    \begin{tabular}{c c c}
        Category & Correct FDSRI & Devices per Category \\
        \hline
        CCTV & 90.83\,\% & 720 \\
        Streaming & 95.24\,\% & 63 \\
        Switch & 100.0\,\% & 4 \\
        Speaker & 100.0\,\% & 3 \\
        Controller & 100.0\,\% & 2 \\
        IP2Serial & 100.0\,\% & 1
    \end{tabular}
    \label{tab:ch9_correct_FDSRI_per_cat}
\end{table}

When looking at the device model categories in Table~\ref{tab:ch9_correct_FDSRI_per_cat}, our SAFER version estimated the FDSRIs for the largest device category (CCTV) with 721 device models achieving a correct FDSRI for 90.84\,\% and the second largest device category (Streaming) achieved a correct FDSRI for 95.16\,\%. 
41 out of 793 device models are estimated to have a higher FDSRI than we observed in real data. This splits into two device models where SAFER predicted the FDSRI to be \emph{medium} but was observed \emph{low}, and 39 device models where SAFER predicted a \emph{high} FDSRI but was observed \emph{medium}. On the contrary, for 53 device models SAFER estimated a too low FDSRI. Those divide in 16 device models with estimated \emph{medium} and observed \emph{high} FDSRI, and 37 device models with estimated \emph{low} and observed \emph{medium} FDSRI. 
The reason being that the PT was previously predicted too high for 29 device models (\emph{medium} instead of \emph{slow}) and for 23 device models, SAFER predicted a too low PT (\emph{slow} instead of \emph{medium}).

We showed that our SAFER version -- implementing the above mentioned mechanisms to predict PT, VT and FDSRI -- is able to indicate the future device security risk indicator for \emph{91.30\,\% of 793} device models correctly. Moreover, we recall that after the initial configuration, SAFER is able to perform device risk scoring in an automated fashion. This enables SAFER to inform embedded-device owners with different security skill-sets about the estimated future device security risks, display technical evidence, e.g.\ discovered software libraries within the firmware, to understand the risks and to support SAFER's users in making informed decisions about possible mitigations to secure the connected network.

\section{Discussion} \label{Discussion}
We point out that we split the data into a training and test-set for predictions of PT and VT. Using this split, we verified that all predictions were compared to observed data. This enabled us to make statements on how accurate the prediction mechanisms perform compared to real, observed data.

\subsection{Limitations}
We assume that publicly registered vulnerabilities for device models and third-party software are verified by other parties prior to registering them at CVE numbering authorities. However, SAFER cannot identify if the device uses, e.g., a vulnerable function in its operation. 
SAFER estimates the device risks based on publicly known vulnerabilities and patch intervals. This does not include unpublished vulnerabilities (zero-days) from, e.g., nation states and underground forums. Thus, SAFER's calculated risk indicator needs to be considered as a risk approximation based on public vulnerability information.
SAFER considers the ``created'' date of a CVE as the date the CVE was publicly registered. SAFER cannot identify if at this date the CVE contained vulnerability information or temporary placeholder information. We assume that when a CVE gets registered for a product, the vendor knows about the vulnerability and is already able to work on patches.
The extended data-set from this work contains data from \axis{} devices, because we considered license statements and release note files as data-sources from this vendor. A wider range of vendors would result in a more realisitc data-set and alleviate possible a possible bias specific to \axis{}.
However, with SAFER, one can create the metrics we discussed for \axis{} for various vendors, e.g., by uploading firmwares to SAFER's firmware analysis component.
\subsection{Comparing with Related Work} \label{discussion:comparing_with_relwork}

The majority of vulnerability prediction approaches require the public availability of source code for all software versions. Massacci et al.~\cite{massacci2010right}, for example, investigated Firefox and compared 18 different vulnerability approaches of related work with their own approach, whereas the 12 best performing ones focus on vulnerability prediction. 
Other vulnerability prediction works require specific amount of past software patches in the source-code as, e.g., \cite{garg2020learning} to train a machine model on. 
Other works are limited to specific programming languages they can parse and process as, e.g., \cite{dam2017automatic}. The last group does not only require source code availability, but also a large code-base as, e.g., \cite{jimenez2016vulnerability} to perform predictions on. 
Unlike related work~\cite{perl2015vccfinder,shin2010evaluating,xiao2020mvp,liu2012software,duan2019automating,xu2017spain} finding vulnerabilities in, e.g., source-code, SAFER uses public sources to fetch known vulnerabilities only. Though, related work~\cite{edkrantz2015predicting,williams2020vulnerability,kudjo2020effect,wu2020vulnerability} with the same approach achieved accurate results. However, we state that by using SAFER's approach to retrieve vulnerabilities, SAFER is not limited in detecting vulnerabilities for, e.g., specific programming languages, but even supports closed-source software and does not require a large source-code base.
SAFER's risk scoring component uses the CVSS standard (in comparison to modified CVSS versions like related work~\cite{wang2011improved,qu2016assessing,wang2018vulnerability,guillen2014risk}) to generate universal and comparable metrics. This is a relevant fact to make device comparisons for different device categories possible.

\section{Summary and Future Work} \label{sec:conclusion}
This paper presented an in-depth evaluation of the FDSRI with 793 device models in comparison to its original work~\cite{oser-tops} with 38 device models and showed that historic vulnerability and patch data can be used to estimate future risks. We used SAFER's risk \emph{prediction \& scoring} component, which aims to automate future security risk assessments by calculating the future device security risk indicator (FDSRI). This indicator identifies how problematic a device might become in the future even if the currently used firmware is considered secure. 

With our enhanced SAFER version, we decreased the vulnerability trend's prediction error by 25.2\,\% and slightly increased the patch trend's prediction error by merely 4.1\,\%. We highlight that in comparison to SAFER's original work~\cite{oser-tops} with 38 device models, we analyzed 793 IoT devices which represents 20.86 times more device models. Hence, our evaluation provides a clear indication of the feasibility but also scalability of SAFER.

SAFER is a suitable tool to establish IoT security awareness in large-scale networks and enables highly automated risk assessments for IoT devices estimating its current as well as future risk. By using our enhanced SAFER version, we correctly estimate the FDSRI for 91.30\,\% of 793 devices.

\subsection{Future Work}
The risk \emph{prediction \& scoring} component could be further enhanced by implementing detection methods for vulnerability chaining, e.g., by using the Common Weakness Enumeration (CWE).

Assuming that SAFER will include more data-sources in the future, we consider evaluating Long Short-Term Memory (LSTM) and further neural network based prediction models for future work which require significantly more data to compute predictions on.

\section*{Acknowledgements}
This work was partially funded by the Sapere Aude: DFF-Starting Grant number 0165-00079B ``Foundations of Privacy Preserving and Accountable Decentralized Protocols''
%
%
\bibliographystyle{splncs04}
\bibliography{bibliography}

\begin{thebibliography}{10}
\providecommand{\url}[1]{\texttt{#1}}
\providecommand{\urlprefix}{URL }
\providecommand{\doi}[1]{https://doi.org/#1}

\bibitem{agarwal2019detecting}
Agarwal, S., Oser, P., Lueders, S.: Detecting iot devices and how they put
  large heterogeneous networks at security risk. Sensors  \textbf{19}(19),
  ~4107 (2019)

\bibitem{bahizad2020risks}
Bahizad, S.: Risks of increase in the iot devices. In: 2020 7th IEEE
  International Conference on Cyber Security and Cloud Computing (CSCloud)/2020
  6th IEEE International Conference on Edge Computing and Scalable Cloud
  (EdgeCom). pp. 178--181. IEEE (2020)

\bibitem{box2011time}
Box, G.E., Jenkins, G.M., Reinsel, G.C.: Time series analysis: forecasting and
  control, vol.~734. John Wiley \& Sons (2011)

\bibitem{chakraborty2021deep}
Chakraborty, S., Krishna, R., Ding, Y., Ray, B.: Deep learning based
  vulnerability detection: Are we there yet. IEEE Transactions on Software
  Engineering  (2021)

\bibitem{dam2017automatic}
Dam, H.K., Tran, T., Pham, T., Ng, S.W., Grundy, J., Ghose, A.: Automatic
  feature learning for vulnerability prediction. arXiv preprint
  arXiv:1708.02368  (2017)

\bibitem{duan2019automating}
Duan, R., Bijlani, A., Ji, Y., Alrawi, O., Xiong, Y., Ike, M., Saltaformaggio,
  B., Lee, W.: Automating patching of vulnerable open-source software versions
  in application binaries. In: NDSS (2019)

\bibitem{duan2021automated}
Duan, X., Ge, M., Le, T.H., Ullah, F., Gao, S., Lu, X., Babar, M.A.: Automated
  security assessment for the internet of things. arXiv preprint
  arXiv:2109.04029  (2021)

\bibitem{edkrantz2015predicting}
Edkrantz, M., Truv{\'e}, S., Said, A.: Predicting vulnerability exploits in the
  wild. In: 2015 IEEE 2nd International Conference on Cyber Security and Cloud
  Computing. pp. 513--514. IEEE (2015)

\bibitem{garg2020learning}
Garg, A., Degiovanni, R., Jimenez, M., Cordy, M., Papadakis, M., Traon, Y.L.:
  Learning to predict vulnerabilities from vulnerability-fixes: A machine
  translation approach. arXiv preprint arXiv:2012.11701  (2020)

\bibitem{guillen2014risk}
Guillen, O.M., Brederlow, R., Ledwa, R., Sigl, G.: Risk management in embedded
  devices using metering applications as example. In: Proceedings of the 9th
  Workshop on Embedded Systems Security. pp.~1--9 (2014)

\bibitem{howell2005median}
Howell, D.C.: Median absolute deviation. Encyclopedia of statistics in
  behavioral science  (2005)

\bibitem{jimenez2016vulnerability}
Jimenez, M., Papadakis, M., Le~Traon, Y.: Vulnerability prediction models: A
  case study on the linux kernel. In: 2016 IEEE 16th International Working
  Conference on Source Code Analysis and Manipulation (SCAM). pp. 1--10. IEEE
  (2016)

\bibitem{johnson2016can}
Johnson, P., Lagerstr{\"o}m, R., Ekstedt, M., Franke, U.: Can the common
  vulnerability scoring system be trusted? a bayesian analysis. IEEE
  Transactions on Dependable and Secure Computing  \textbf{15}(6),  1002--1015
  (2016)

\bibitem{josang-subjective-2016}
J{\o}sang, A.: Subjective Logic: A Formalism for Reasoning Under Uncertainty.
  Artificial Intelligence: Foundations, Theory and Algorithms, Springer
  International Publishing Switzerland (2016). \doi{10.1007/978-3-319-42337-1}

\bibitem{kudjo2020effect}
Kudjo, P.K., Chen, J., Mensah, S., Amankwah, R., Kudjo, C.: The effect of
  bellwether analysis on software vulnerability severity prediction models.
  Software Quality Journal  \textbf{28}(4),  1413--1446 (2020)

\bibitem{le2018security}
Le, N.T., Hoang, D.B.: Security threat probability computation using markov
  chain and common vulnerability scoring system. In: 2018 28th International
  Telecommunication Networks and Applications Conference (ITNAC). pp.~1--6.
  IEEE (2018)

\bibitem{li2021understanding}
Li, Q., Tan, D., Ge, X., Wang, H., Li, Z., Liu, J.: Understanding security
  risks of embedded devices through fine-grained firmware fingerprinting. IEEE
  Transactions on Dependable and Secure Computing  (2021)

\bibitem{li2021sysevr}
Li, Z., Zou, D., Xu, S., Jin, H., Zhu, Y., Chen, Z.: Sysevr: A framework for
  using deep learning to detect software vulnerabilities. IEEE Transactions on
  Dependable and Secure Computing  (2021)

\bibitem{li2018vuldeepecker}
Li, Z., Zou, D., Xu, S., Ou, X., Jin, H., Wang, S., Deng, Z., Zhong, Y.:
  Vuldeepecker: A deep learning-based system for vulnerability detection. arXiv
  preprint arXiv:1801.01681  (2018)

\bibitem{liu2012software}
Liu, B., Shi, L., Cai, Z., Li, M.: Software vulnerability discovery techniques:
  A survey. In: 2012 fourth international conference on multimedia information
  networking and security. pp. 152--156. IEEE (2012)

\bibitem{massacci2010right}
Massacci, F., Nguyen, V.H.: Which is the right source for vulnerability
  studies? an empirical analysis on mozilla firefox. In: Proceedings of the 6th
  International Workshop on Security Measurements and Metrics. pp.~1--8 (2010)

\bibitem{oser-frontend}
Oser, P., Feger, S., Wo{\'z}niak, P.W., Karolus, J., Spagnuelo, D., Gupta, A.,
  L{\"u}ders, S., Schmidt, A., Kargl, F.: Safer: Development and evaluation of
  an iot device risk assessment framework in a multinational organization.
  Proceedings of the ACM on Interactive, Mobile, Wearable and Ubiquitous
  Technologies  \textbf{4}(3),  1--22 (2020)

\bibitem{oser-tops}
Oser, P., van~der Heijden, R.W., L{\"u}ders, S., Kargl, F.: Risk prediction of
  iot devices based on vulnerability analysis. ACM Transactions on Privacy and
  Security  \textbf{25}(2),  1--36 (2022)

\bibitem{oser-CCD}
Oser, P., Kargl, F., L{\"u}ders, S.: Identifying devices of the internet of
  things using machine learning on clock characteristics. In: International
  Conference on Security, Privacy and Anonymity in Computation, Communication
  and Storage. pp. 417--427. Springer (2018)

\bibitem{perl2015vccfinder}
Perl, H., Dechand, S., Smith, M., Arp, D., Yamaguchi, F., Rieck, K., Fahl, S.,
  Acar, Y.: Vccfinder: Finding potential vulnerabilities in open-source
  projects to assist code audits. In: Proceedings of the 22nd ACM SIGSAC
  Conference on Computer and Communications Security. pp. 426--437 (2015)

\bibitem{qu2016assessing}
Qu, Y., Chan, P.: Assessing vulnerabilities in bluetooth low energy (ble)
  wireless network based iot systems. In: 2016 IEEE 2nd International
  Conference on Big Data Security on Cloud (BigDataSecurity), IEEE
  International Conference on High Performance and Smart Computing (HPSC), and
  IEEE International Conference on Intelligent Data and Security (IDS). pp.
  42--48. IEEE (2016)

\bibitem{rodriguezsuperspreaders}
Rodr{\'\i}guez, E., Noroozian, A., van Eeten, M., Ga{\~n}{\'a}n, C.:
  Superspreaders: Quantifying the role of iot manufacturers in device
  infections

\bibitem{russell2018automated}
Russell, R., Kim, L., Hamilton, L., Lazovich, T., Harer, J., Ozdemir, O.,
  Ellingwood, P., McConley, M.: Automated vulnerability detection in source
  code using deep representation learning. In: 2018 17th IEEE international
  conference on machine learning and applications (ICMLA). pp. 757--762. IEEE
  (2018)

\bibitem{shin2010evaluating}
Shin, Y., Meneely, A., Williams, L., Osborne, J.A.: Evaluating complexity, code
  churn, and developer activity metrics as indicators of software
  vulnerabilities. IEEE transactions on software engineering  \textbf{37}(6),
  772--787 (2010)

\bibitem{shivraj2017graph}
Shivraj, V., Rajan, M., Balamuralidhar, P.: A graph theory based generic risk
  assessment framework for internet of things (iot). In: 2017 IEEE
  International Conference on Advanced Networks and Telecommunications Systems
  (ANTS). pp.~1--6. IEEE (2017)

\bibitem{taylor2018forecasting}
Taylor, S.J., Letham, B.: Forecasting at scale. The American Statistician
  \textbf{72}(1),  37--45 (2018)

\bibitem{vilches2018towards}
Vilches, V.M., Gil-Uriarte, E., Ugarte, I.Z., Mendia, G.O., Pis{\'o}n, R.I.,
  Kirschgens, L.A., Calvo, A.B., Cordero, A.H., Apa, L., Cerrudo, C.: Towards
  an open standard for assessing the severity of robot security
  vulnerabilities, the robot vulnerability scoring system (rvss). arXiv
  preprint arXiv:1807.10357  (2018)

\bibitem{wang2018vulnerability}
Wang, H., Chen, Z., Zhao, J., Di, X., Liu, D.: A vulnerability assessment
  method in industrial internet of things based on attack graph and maximum
  flow. Ieee Access  \textbf{6},  8599--8609 (2018)

\bibitem{wang2011improved}
Wang, R., Gao, L., Sun, Q., Sun, D.: An improved cvss-based vulnerability
  scoring mechanism. In: 2011 Third International Conference on Multimedia
  Information Networking and Security. pp. 352--355. IEEE (2011)

\bibitem{williams2020vulnerability}
Williams, M.A., Barranco, R.C., Naim, S.M., Dey, S., Hossain, M.S., Akbar, M.:
  A vulnerability analysis and prediction framework. Computers \& Security
  \textbf{92},  101751 (2020)

\bibitem{wu2020vulnerability}
Wu, S., Wang, C., Zeng, J., Wu, C.: Vulnerability time series prediction based
  on multivariable lstm. In: 2020 IEEE 14th International Conference on
  Anti-counterfeiting, Security, and Identification (ASID). pp. 185--190. IEEE

\bibitem{xiao2020mvp}
Xiao, Y., Chen, B., Yu, C., Xu, Z., Yuan, Z., Li, F., Liu, B., Liu, Y., Huo,
  W., Zou, W., et~al.: $\{$MVP$\}$: Detecting vulnerabilities using
  patch-enhanced vulnerability signatures. In: 29th $\{$USENIX$\}$ Security
  Symposium ($\{$USENIX$\}$ Security 20). pp. 1165--1182 (2020)

\bibitem{xu2017spain}
Xu, Z., Chen, B., Chandramohan, M., Liu, Y., Song, F.: Spain: security patch
  analysis for binaries towards understanding the pain and pills. In: 2017
  IEEE/ACM 39th International Conference on Software Engineering (ICSE). pp.
  462--472. IEEE (2017)

\bibitem{zhou2019devign}
Zhou, Y., Liu, S., Siow, J., Du, X., Liu, Y.: Devign: Effective vulnerability
  identification by learning comprehensive program semantics via graph neural
  networks. arXiv preprint arXiv:1909.03496  (2019)

\end{thebibliography}
\end{document}